# Intra-instrument channel workable, optical-resolution photoacoustic and ultrasonic mini-probe system for clinical gastrointestinal endoscopy

**Minjae Kim,[1,4] Kang Won Lee,[2,4] KiSik Kim,[1] Oleksandra Gulenko,[1] Cheol Lee,[3] Bora Keum,[2] Hoon Jai Chun,[2] Hyuk Soon Choi,[2,5] Chae Un Kim,[3,6] and Joon-Mo Yang[1,7]**

[1]*Center for Photoacoustic Medical Instruments, Department of Biomedical Engineering, Ulsan National Institute of Science and Technology (UNIST), Ulsan 44919, South Korea*
[2]*Division of Gastroenterology and Hepatology, Department of Internal Medicine, Korea University College of Medicine, Seoul 02841, South Korea*
[3]*Department of Physics, UNIST, Ulsan 44919, South Korea*
[4]*These authors contributed equally to this work*
[5]*mdkorea@gmail.com*
[6]*cukim@unist.ac.kr*
[7]*jmyang@unist.ac.kr*

**Abstract:** There has been a long-standing expectation that the optical-resolution embodiment of photoacoustic tomography could have a substantial impact on gastrointestinal endoscopy by enabling microscopic visualization of the vasculature based on the endogenous contrast mechanism [Med. Phys. 35, 5758 (2008)]. Although multiple studies have demonstrated the in vivo imaging capability of a developed imaging device over the last decade, the implementation of such an endoscopic system that can be applied immediately when necessary via the instrument channel of a video endoscope has been a challenge. In this study, we developed a 3.38-mm diameter catheter-based, integrated optical-resolution photoacoustic and ultrasonic mini-probe system and successfully demonstrated its intra-instrument channel workability for the standard 3.7-mm diameter instrument channel of a clinical video endoscope based on a swine model. Through the instrument channel, we acquired the first in vivo photoacoustic and ultrasonic dual-mode endoscopic images from the esophagogastric junction, which is one of the most frequently cited anatomical sites in the gastrointestinal tract in relation to Barrett's esophagus. Moreover, from a rat colorectum in vivo imaging experiment, we visualized hierarchically developed mesh-like capillary networks with a hole size as small as ~50 μm, which suggests the potential level of image details that could be photoacoustically provided in clinical settings in the future.

## 1. Introduction

One of important technical aspects of photoacoustic (PA) tomography (PAT) in terms of clinical application is that related imaging systems can be embodied in a form that is integrated with the conventional ultrasound (US) imaging function [1–8]. When pursing the clinical application of PAT, the integrability is an attractive aspect, as the US imaging technique has played an indispensable role in numerous clinical areas and also the field of application of the technique is still being broadened, even in minimally-invasive imaging in forms such as endoscopic ultrasound (EUS) and intravascular ultrasound (IVUS) [9,10]. Thus, if a miniaturized PA imaging device developed for gastrointestinal (GI) endoscopy also possesses the EUS function, it could complement the current limitations of EUS based on the superior microvascular angiographic and functional imaging capability of PAT [1–8].

After the first idea of applying PAT to endoscopy in 1996 [11], a variety of device concepts have been proposed to translate its benefits into clinical settings [12–39]. Among these, the following are notable device concepts that are potentially applicable for GI endoscopy. The first is the application of a built-in scanning mechanism in which a micro-motor is placed at the scanning tip to function as an actuator [15–18]. The key benefit of this mechanism is that it enables easy embodiment of necessary optical and electrical coupling in the proximal part of the probe, as associated optical and electrical elements can be statically connected to peripheral elements. However, due to the accompanying technical challenge in implementing the key mechanical unit, such an endoscopic probe with a sufficiently short rigid distal section that can freely pass through the instrument channel of a video endoscope has not been demonstrated thus far. To avoid such a technical challenge, a torque coil-based proximal actuation mechanism was introduced in 2014 [22] and has become the dominant mechanism for most follow-up mini-probes for PA endoscopy (PAE) [19–21] or catheter probes for intravascular photoacoustic (IVPA) imaging application [23–35]. Although it is expected that it will take a considerable amount of time for commercialized devices based on the related mechanism to be used for real human subjects, one of the key strengths of this probe is a flexible body along with a short rigid distal section. These two mechanisms are based on single-element transducer-based mechanical scanning. However, since 1998, pure optics-based US detection mechanisms have also been investigated because they exhibit an exceptional sensitivity [36]. Among related embodiments [36–40], the result reported by Ansari et al. in 2018 demonstrated a 50,000-channel forward-viewing probe concept based on a Fabry–Pérot interferometer [39]. Although the probe was embodied in a rigid form, they achieved the vastest number of array channels, which was never possible with the conventional piezoelectric method.

In this study, by applying the conventional piezoelectric material-based US detection mechanism and the torque coil-based proximal actuation mechanism, we developed a 3.38-mm diameter integrated PAE-EUS mini-probe system, which can be applied immediately when a suspicious tissue is found during a video endoscopic procedure. Although a certain endoscopic application would require a high sensitivity of the optical US detection mechanism, particularly in the case where the applicable optical energy is limited, we considered the traditional piezoelectric material-based US detection mechanism still valuable and important because it offers the irresistible benefit of enabling seamless integration with the EUS technology, thereby accelerating the clinical translation of PAE based on the established platforms of EUS that clinicians are accustomed to. Moreover, we adopted the torque coil-based proximal actuation mechanism because it is a typical mechanism that is applied to current clinical EUS mini-probes [9,10].

Of course, most of the IVPA devices reported in recent years have been implemented based on the same mechanism [26–35], and one of the reports even achieved the image of a swine coronary artery in vivo [28]. However, none of them has successfully demonstrated the imaging of the upper GI tract of a large animal via the instrument channel of a clinical video endoscope; only imaging of a rat colorectum has been reported with regard to GI tract imaging [19,20]. Moreover, the PA image qualities have been merely limited to acoustic-resolution (AR) imaging because the devices relied on multi-mode optical fibers with a large core diameter (>100 μm) (note that this is one of the important factors determining the transverse resolution of an implemented device). In other words, although the torque coil-based proximal actuation mechanism could be commonly applied to an IVPA and GI endoscopic probe, achieving the latter embodiment is never minor; this is because it requires solving an additional technical issue that involves completely sealing the probe sheath near the proximal optical and electrical rotary junction as well as at the distal tip to completely isolate the acoustic matching medium filled inside. Furthermore, in order to achieve a higher spatial resolution, it is essential to embody an optical rotary junction based on an optical fiber with a narrower core diameter.

In this study, we overcame the technical limits of previous embodiments and achieved optical-resolution (OR) PA imaging as low as ~13 μm even in the presented in vivo imaging demonstrations by implementing a customized optical rotary junction and distal optics. In this article, we describe our engineering strategies formulated to surpass existing imaging performances and also present related embodiment results of newly addressed technical issues, such as those pertaining to the coupling of the electric signal near the rotary junction and signal balancing between the PA and US modes when sharing the same digitizer, both of which were successfully proven through the presented in vivo imaging results.

## 2. Results

## A. PAE-EUS mini-probe system

### 1. System composition and distal structure of the mini-probe

Figure 1(a) illustrates an overall view of the lab-made mini-probe system constructed on a movable cart for its easy transportation in an endoscopy room. The system consists of a mini-probe, which is a key module designed to be connectable to a driving unit in an interchangeable manner and has a flexible section with a length of ~1.8 m and an outer diameter (OD) of 3.38 mm, including the tubing dimension; a driving unit that drives the mini-probe; and their peripheral systems including a data acquisition (DAQ) computer, laser controller, and DC power supply.

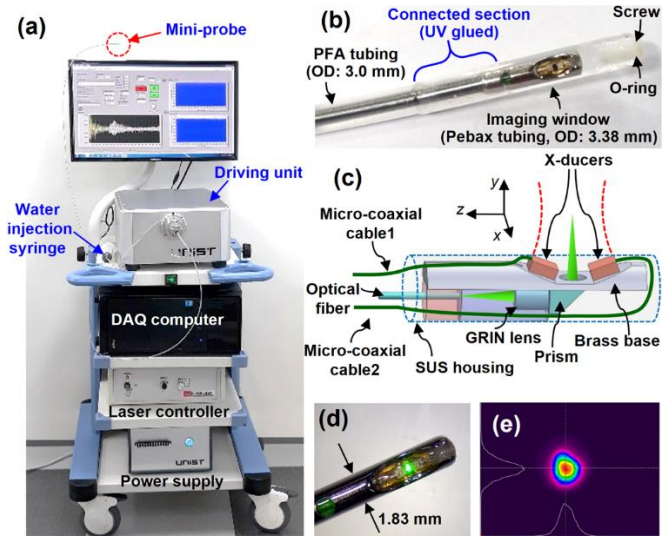

Fig. 1. Integrated PAE-EUS mini-probe system and structure of the distal section of the mini-probe. (a) Photo of the entire system. (b, c) Photo and schematic of the distal section. Note that the Cartesian coordinate system denoted by (x, y, z) is a moving and rotating coordinate system along with the scanning tip. (d) Photo of the scanning tip emitting laser beam. (e) 2D beam profile of the laser beam at the optical focus.

Figures 1(b) and 1(c), respectively, depict a photo and corresponding schematic of the mini-probe's distal section where rotational scanning is performed. The distal section encapsulated by a stainless steel (SUS) tube had a length of 10 mm and OD of 1.83 mm; moreover, the distal section was connected to a 1.8-m long torque coil with a 1.3-mm OD and 0.7-mm inner diameter (ID), in which a multi-mode optical fiber (FG010LDA, 10-μm core dia., Thorlabs, USA) and two micro-coaxial cables from two transducers (40 MHz, PZT composite, 0.6 mm × 0.5 mm × 0.2 mm, Blatek, USA) were placed.

As with conventional EUS mini-probes [9,10], the distal and torque coil sections must be properly sealed in a sheath (i.e., tubing or catheter) to avoid their direct contact with a target tissue during their fast rotation. For the developed mini-probe, the OD and ID of a necessary sheath could be appropriately selected depending on the target application. However, for this study, we created a sheath in our lab by combining two different types of commercial tubing—that is, Perfluoroalkoxy (PFA) for the 1.8-m long torque coil section (OD/ID: 3.0/2.0 mm) and Pebax 7033 for the imaging window (OD/ID: 3.38/3.17 mm, custom ordered, A.P. Extrusion, Inc., USA) [Fig. 1(b)]; this was done because of the non-availability of such a one-body sheath with two different wall thicknesses among currently available commercial tubing line-ups. We chose the two materials because PFA exhibits low mechanical friction, which is good for the smooth rotation of the torque coil, whereas Pebax 7033 has high optical and acoustic transparency. Moreover, applying a tubing with such a thin wall (i.e., ~100 μm) for the imaging window section was necessary to minimize the distortion of a focused laser beam.

Considering the 3.38-mm OD of the imaging window, we targeted a working distance (WD) of 1.46 mm in water, which was defined by the distance from the surface of the SUS tubular housing with the 1.83-mm OD to the optical focus created by a gradient-index (GRIN) lens (0.5-mm diameter, ~0.1 pitch; altered from #64-515, Edmund Optics, USA). The physical length of the GRIN lens was reduced to 0.44 mm from its original length of 1.15 mm by an in-house polishing process; this was done in keeping with a previously reported technical note that recommended an avoidance of direct contact of the GRIN lens to the tip of the optical fiber [18], while achieving the targeted WD. Consequently, the laser beam diverging from the optical fiber tip was propagated in air by a ~1.1-mm interval, converged by the GRIN lens, and finally focused after being reflected at the hypotenuse of a right-angle prism (#66-772, Edmund Optics, USA) by the total internal reflection (TIR).

To detect PA waves effectively and also enable co-registered US imaging without the issue of misalignment between the optical and acoustic axes, we applied a weak US-focusing strategy based on the dual transducers. As illustrated in Fig. 1(c), we installed the two transducers onto the two inclined surfaces with a 17° angle to the z-axis and a separation of ~0.5 mm from each other so that the focused laser beam could come out through the gap between. In terms of securing the optical TIR, it was important that one leg of the prism be affixed around the 0.5-mm diameter hole of a brass base [Fig. 1(c)], so that it could also function as a sealing point that isolates the associated optics from the acoustic matching liquid filled in the tubing and not merely function as an optical aperture.

After the probe implementation, we measured the laser beam diameter at the optical focus by using a beam profiler (SP620U, OPHIR Beam Gauge). Figures 1(d) and 1(e), respectively, depict the microscope image of the scanning tip that emits a laser beam and a 2D beam profiler image acquired at the focal plane, from which the FWHM beam diameter was determined to be ~8 μm in air. Thereafter, we performed an optical WD measurement experiment by imaging a surgical blade submerged in water according to the procedure described in Fig. S1 of Supplement 1. Through the experiment, we found that although we targeted an optical WD of 1.46 mm in water, the actual measurement value was ~0.89 mm. In addition, we determined the PA/US image resolutions to be ~17–67 μm/191–355 μm for the transverse direction (x-direction) and ~61–130 μm/66–75 μm for the radial direction (i.e., y-direction) (see Fig. 1(c) for the axis definition).

Although the resolutions determined by the above method revealed a high radial distance dependence and the best PA transverse resolution appeared to be just 17 μm—which is two times greater than the measured optical beam diameter—we were able to validate that the PA transverse resolution of the implemented probe was as low as ~13 μm, while the optical WD was determined to be slightly differently (~0.86 mm) from the above value 0.89 mm when we re-estimated them through the analysis presented in Fig. S5 of Supplement 1. Based on the two determined WD values, we assume that during the rotational scanning of the scanning tip, the focal spot of the laser beam traces a circle of diameter ~3.54 mm–3.6 mm, which is slightly larger than the OD of the imaging window. We attribute the close-shift of the WD to the imperfection of our assembling work for the distal optics.

Unlike an IVUS imaging probe [9,10], whose intraluminal acoustic matching is usually achieved by the continuous injection of a saline solution during the related imaging procedure, EUS mini-probes for GI endoscopy are usually manufactured in a form that requires the acoustic matching medium to be permanently filled inside it. Although such a permanent filling was also desirable for the developed mini-probe, we had to inevitably apply the previously reported temporal water injection strategy [18] because there was still a concern regarding the possible degradation of optical and acoustic performance caused by the impurities involved during the in-house assembling process. By using the water injection syringe presented in Fig. 1(a), we filled the inner space of the probe with deionized water from the proximal part to the distal end of the probe and finally sealed the distal tip using a plastic screw with a silicon O-ring [Fig. 1(b)]. After an experiment was conducted, the injected water was drained out due to the abovementioned concern.

## 2. Driving unit

To drive the mini-probe, we designed and implemented a driving unit as depicted in Fig. 2(a), in which an oscillator and delay generator unit, pulse amplifier, RF switch and signal amplifier unit, clamping nozzle, stepping motor, and laser head were integrated. Notably, the proximal part of the mini-probe (i.e., bearing unit) [Fig. 2(b)] was designed to be clamped into the clamping nozzle of the driving unit [Fig. 2(a)], and then the timing pulley and SMA connector formed around the bearing unit were engaged and connected to the stepping motor and the RF switch, respectively, to transfer the necessary mechanical torque as well as associated electric signals.

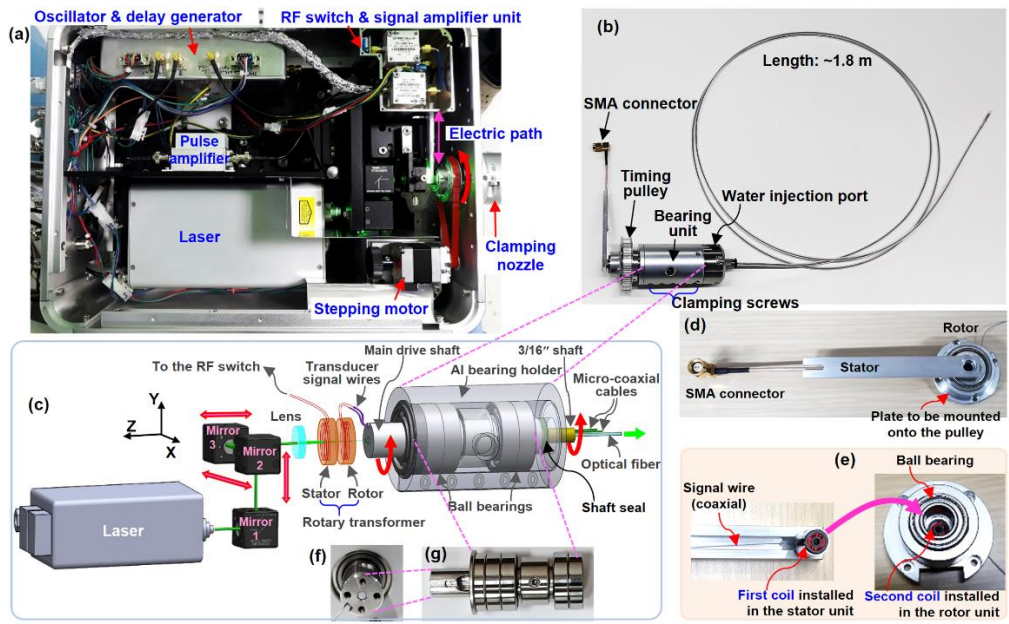

Fig. 2. Driving unit. (a) Photo depicting the interior of the driving unit. (b) Entire view of the mini-probe. (c) Schematic presenting the configuration of the integrated optical and electrical rotary junction along with the laser beam path. Note that the Cartesian coordinate system denoted by (X, Y, Z) is a moving but non-rotating coordinate system introduced to translationally trace the position of the center of the scanning tip. Photos depicting the entire view (d) and internal structure (e) of the rotary transformer unit.

One of the key achievements of this study compared to previous studies [19–35] is that we applied a rotary transformer concept for the first time to communicate electric signals for PA and US imaging at the associated electric rotary joint, thereby successfully acquiring the required images, which are presented later in the paper. The demonstrated coupling mechanism could be potentially useful because, compared to the slip ring-based signal coupling mechanism that was typically applied in previous studies [19–35], the rotary transformer-based signal coupling mechanism does not involve any contact resistance fluctuation and transmission efficiency degradation caused by aging; moreover, it also has an advantage in terms of unit cost. In addition, it can also play the role of an electric impedance matcher between the transducers and amplifier and that of a DC blocker as well.

A rotary transformer unit created on the basis of a set of pot cores was installed inside the timing pulley depicted in Fig. 2(b), and its approximate configuration with respect to coupling optics is depicted in Fig. 2(c). Furthermore, as illustrated in Figs. 2(d) and 2(e), the rotary transformer unit consisted of the rotor and stator, and associated signals were transferred between them through a pair of coils (turns ratio: 1:1) that were located within the rotor and stator, respectively. A ball bearing installed in the rotor unit [Fig. 2(e)] provided a smooth rotational condition, which enabled a stable signal transfer while the rotor and the torque coil rotate. In addition, we measured one-way electric signal transmission efficiency of the rotary transformer according to the method presented in Fig. S2 of Supplement 1 and observed an efficiency as high as ~95% for the engagement of a 50-Ω load. As depicted in Fig. 2(c), the two micro-coaxial cables from the two transducers were combined inside the bearing unit and connected to the second coil included in the rotor [Fig. 2(e)], which resulted in an electrically parallel connection effect to the employed transducers.

In addition to the electric characteristics, the two ferrite cores of the rotary transformer unit also featured a 1.8-mm diameter hole formed at the center [Fig. 2(e)]; thus, the laser beam could be injected into the entrance tip of an optical fiber via the holes, as depicted in Fig. 2(c), thereby constituting an integrated optical and electrical rotary junction. The laser beam emitted from the laser system (532 nm, ~2 ns pulse length, SPOT 10-200-532, Elforlight) was guided by three turning prism mirrors (CCM5-E02/M, Thorlabs), whose positions were translationally adjustable along the X-Y-Z directions and focused on the optical fiber by a focusing lens ($f$ = 30 mm, LA1289-A, Thorlabs). Figure 2(f) is an en face view of the custom-ordered main drive shaft (OD: 12 mm, SUS) at the center of which the optical fiber that extended up to the scanning tip was placed accurately with the aid of a ceramic optical fiber ferrule (CF126-10, Thorlabs).

In creating such an optical rotatory junction, it is key to provide a stable rotational condition to the optical fiber rotating with a torque coil. To better achieve this, we employed five ball bearings, including four high precision angular ball bearings [Fig. 2(g)], which were modularized in an aluminum (Al) bearing holder through a clapping method [Fig. 2(b)]—this was the simplest method that minimized the runout of the optical fiber's rotational axis during its operation. However, considering the possible imperfection of our assembling work, we employed an optical fiber with a 10-µm core diameter and also applied a broad injection strategy of the laser beam by employing a plano-convex lens with a low NA (LA1289-A, Thorlabs)—the FWHM beam diameter measured at its focal plane was ~20 µm.

Unlike the previous IVPA probes whose device workability was demonstrated based on a temporal injection of a saline solution [28,33,34], creating a PA probe whose inner tubular space could be filled with an appropriate acoustic matching medium for a long time was a key aspect in successfully applying it to GI endoscopy via the instrument channel of a commercial video endoscope. To resolve this issue, we employed another 3/16″-diameter SUS shaft that was operated through installation in the 12-mm diameter main drive shaft, while providing the necessary mechanical sealing for the injected deionized water in conjunction with the "shaft seal" [see Fig. 2(c)] during its rapid rotation.

Based on the presented hardware, we performed co-registered PA and US imaging at an A-line acquisition rate of 16 kHz for each imaging mode, which was limited by the maximum pulse repetition rate (PRR) of the employed laser system (the operation sequence is described in the next section). For the presented in vivo imaging experiments, the laser pulse energy did not exceed ~1 µJ because it was a maximum energy attainable by the employed laser system when measured at the optical output (note that the applied pulse energy is approximately two times greater than that in a previous study) [18].

## 3. Circuitry and operation sequence for co-registered PA and US dual-mode imaging

Co-registered dual-mode PA and US imaging based on a commercial US pulser/receiver has already been demonstrated in previous studies, along with the interleaving concept of required laser and electric pulses and detecting corresponding responses in turn [16,17,19]. In this study, however, we achieved dual-mode imaging by implementing a customized circuit. A technical issue when applying a commercially available US pulser/receiver is that it is fundamentally difficult to make the signal levels of the two imaging modes comparable to each other because the available electric pulse energy levels of a commercial pulser are limited. In other words, although the lowest possible pulse energy is applied, the signal amplitudes of the US mode are usually much greater than those of the PA mode because PA waves are induced by laser pulses and, thus, subjected to deposited laser pulse energy (note that in the cases of the pulser/receiver models utilized in Refs. 22–35, an amplitude of ~80 V is the minimum voltage value available for the acoustic pulsing). A simple option to solve such an issue would be to add an attenuator between a pulser/receiver and an engaged transducer because then the magnitude of the electric pulse applied to transducer could be attenuated [16]. However, this method inevitably has the major drawback of attenuating upcoming PA and US signals as well.

Recognizing this technical issue, we implemented a first version of a customized pulser/receiver circuit by employing a commercial RF switch (ZASWA2-50DR-FA+, Mini-Circuits, USA) and applying the high-voltage pulse amplifier concept described in Reference 41. Figure 3(a) depicts an approximate block diagram of the implemented circuit, and Figs. 3(b) and 3(c) represent the timing charts of the trigger pulses applied to the associated key elements.

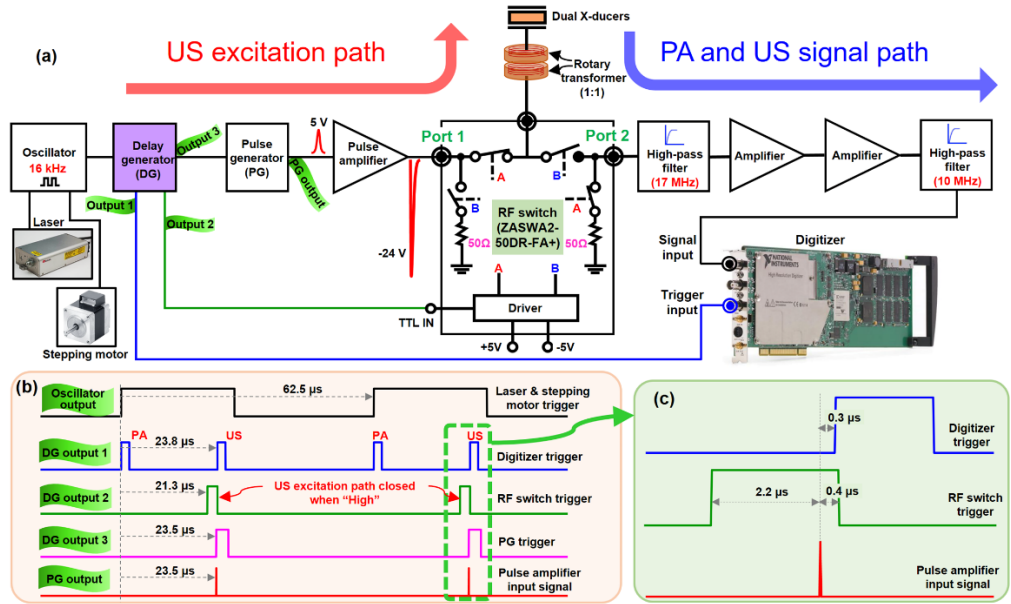

Fig. 3. Circuitry of the PAE-EUS mini-probe system and its operation sequence. (a) Block diagram depicting the electric connection of key elements. (b) Digital timing diagrams generated at the marked positions in (a). (c) Magnified view of the dashed area depicted in (b).

First, square-shaped pulses with a 16-kHz frequency (T = 62.5 μs) are generated by an oscillator that is utilized as a fundamental frequency at which all the sub-electronic and peripheral systems are synchronized, including the stepping motor and laser system. Then, the pulses are sent to a lab-made delay generator (DG), in which the pulses are reshaped into three different types of waveforms with a preset time delay and pulse duration in order to trigger associated electronic elements. Among the three waveforms, the trigger waveform sent to the digitizer (i.e., DG output 1 illustrated in Fig. 3(b)) can be understood as a reference timeline, according to which the dual-mode PAE and EUS imaging sequence and associated DAQ events take place. As illustrated, for every single pulse input to the DG, dual pulses are created with a time lag of 23.8 μs, which respectively correspond to a trigger pulse for PA and US signal acquisitions.

At the rising edge of the first pulse (see DG output 1 depicted in Fig. 3(b)), a laser pulse is fired from the laser. Then, subsequent PA signals detected by the dual transducers are transferred up to the digitizer along the blue-arrow path because, for 21.3 μs from the first pulse, all the internal switches denoted by "B" in the RF switch are in a "closed state," as the voltage level of the waveform that drives the RF switch (i.e., DG output 2) is in a low state. However, near the rising edge of the second pulse, more complex switching events occur throughout the associated electronic elements. First, through Output 3 of DG, a trigger pulse (see the corresponding timing chart of DG output 3) is sent to a pulse generator (PG) to activate it, thereby immediately generating a sharp pulse with an amplitude of 5V, as illustrated in Figs. 3(a) and 3(b). Thereafter, the output pulse of PG is sent to an inverting pulse amplifier to be amplified up to 24 V (~20 ns duration); finally, the output pulse is sent to the transducers via the RF switch [Fig. 3(a)]. Of course, at this moment, all the internal switches denoted by "A" in the RF switch are in a "closed state," thus creating a necessary electric route, as the voltage level of the waveform that drives the RF switch is in a high state.

In brief, as soon as the set acoustic energy is released through the transducers, the internal switches of the RF switch are changed immediately to a "signal receiving mode" because the voltage level of the waveform that drives the RF switch drops to a low state just after 0.4 μs from the US transducer excitation moment [see Fig. 3(c)]. In other words, the US excitation path depicted by a bold red arrow in Fig. 3(a) is closed only when an excitation pulse of 24V is sent to the transducers for pulsing. Figure 3(c) presents a magnified view of the dashed box included in Fig. 3(b), from which one can better differentiate the relative locations of the trigger waveforms applied to the digitizer and the RF switch with respect to the US pulse firing moment. The amplitude of 24V set for the transducer excitation was a voltage value that was much lower than that of a commercial US pulser/receiver, which typically exceeds 80V. Although the implemented pulse amplifier circuit could also produce such a high voltage output, as it was merely determined by an applied Vcc, we chose the amplitude of 24V for the excitation because it was empirically found that just around the set excitation voltage, the signal level of the US mode was slightly lower than that of the PA mode—that is, the full dynamic range of the DAQ was adapted to the PA imaging mode.

Of course, the explained sequences take place continuously, while the scanning tip actuated by the stepping motor continuously rotates in accordance with the fundamental frequency. Thus, the 16-kHz fundamental frequency is also the A-line acquisition rate of the endoscopic system. However, considering the available PRR of the employed laser system, we set the number of A-lines for one full PA B-scan to be 800, which results in a 20-Hz B-scan frame rate for each imaging mode; this was done to maximize the rotational scanning speed even though we could not fully harness the 8-μm level optical resolving power of the PA imaging mode. Then, the PA or US signals detected by the dual-element US transducers are amplified using dual amplifiers (ZFL-500LN+, Mini-Circuits), recorded by the digitizer (200 MHz, 12 bits, PCI-5124, National Instruments, USA), and finally displayed on a monitor screen in real time, as depicted in Fig. 1(a). However, in the monitor display of the current study, we presented the acquired images in a rectangular format because if we applied a necessary coordinate transform, intermittent missing of image slices occurred at the set acquisition rate of 16 kHz (note that if we set 400 data points for the PA mode, double the data points must be recorded for the US mode to cover the same imaging depth). More details on the circuitry of the implemented mini-probe and driving unit are presented in Fig. S3 of Supplement 1.

**B. Rat colorectum 3D imaging in vivo**

In order to test the system's in vivo imaging capability, we imaged the colorectum of a Sprague Dawley rat (~450 g; OrientBio, Seongnam, S. Korea) using the setup depicted in Fig. 4(a). As the descending colon of a rat typically does not involve serious motion artifacts, it was possible to acquire a 3D in vivo data set by adding a pulling back translation motion to the mini-probe, which resulted in creating a helicoidal scan plane [Fig. 4(b)] for the scanning tip that constantly rotated inside the tubing [Fig. 4(c)]. In the experiment, considering a possible large variation of a target distance, we set the A-line data length sufficiently at as high as 400 points at a sampling rate of 200 MHz and acquired over 4000 B-scan slices for each imaging mode (~3.6 GB) from a C-scan performed at a pullback pitch of 16 μm and speed of ~0.32 mm/s. Of course, once a C-scan was initiated, A-line signals were continuously transferred and recorded into the DAQ computer at 16 kHz. However, we separated every 800 A-lines and regarded them as one B-scan for the purpose of real-time computer display and recording.

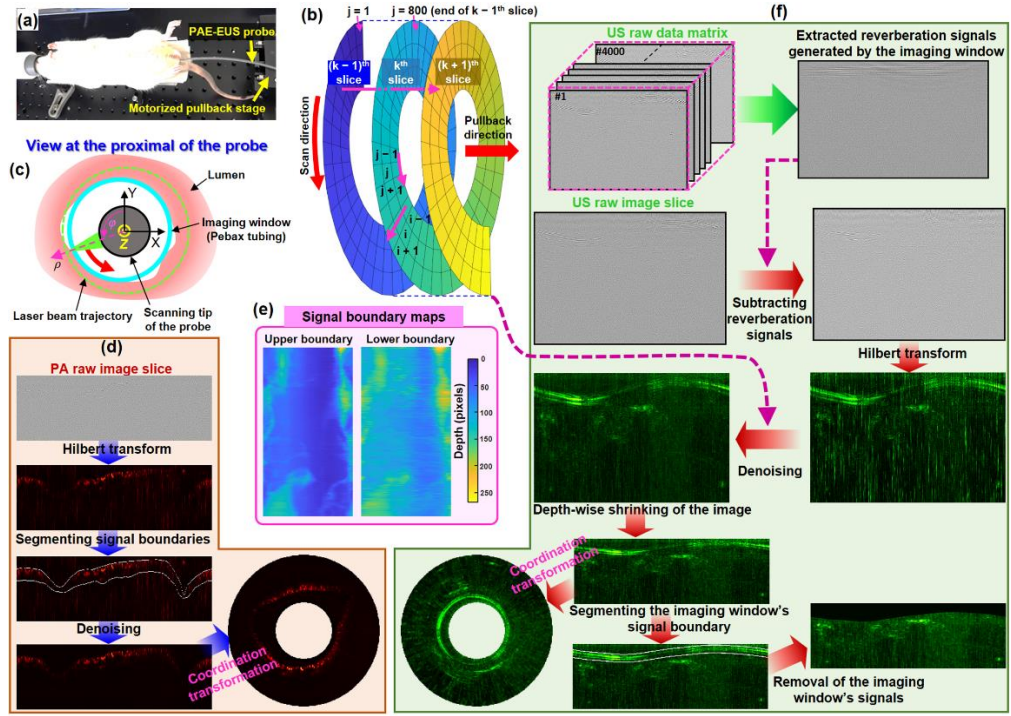

Fig. 4. Rat colorectum imaging experimental setup, coordinate definition, and PAE-EUS image processing procedures. (a) Photo depicting the rat colorectum imaging experiment. (b) Schematic illustrating the scanning trajectory of the imaging probe and the index definition to discriminate the data points acquired from the scanning trajectory. (c) Coordinate definition to represent the rotation direction and angle of the scanning tip. (d) PA image processing procedure. (e) Signal boundary maps, extracted from the PA image data set acquired from a rat colorectum, according to the image segmentation procedure illustrated in (d). (f) US image processing procedure.

Basically, the signal-to-noise ratio (SNR) of in vivo PA data was greater than 30 dB when it was compared to the thermal noise. However, there were substantial effects caused by the interference noise that occurred due to the imperfect shielding to the associated electronic hardware. In a raw B-scan image, interference noise-affected pixels appeared randomly in their positions, but appeared in a rain-like pattern. This was because a range of data points in an A-line were serially affected and, thus, it was not difficult to ascertain whether or not a pixel was affected by an interference noise based on a visual judgement. Prior to presenting the in vivo imaging results, we present an approximate explanation for our interference noise elimination algorithm.

Assuming that either a PA or US C-scan data set was saved in a 3D matrix—for example $C(i, j, k)$, where "i" denotes a radial index ranging from 1 to 400, "j" the step number ranging from 1 to 800, and "k" a slice number ranging from 1 to 4000, we eliminated the interference noise in accordance with two different procedures for the two imaging modes described below.

First, for each PA raw image slice [Fig. 4(d)], we applied the Hilbert-transform to extract the envelope of the bipolar signal. Then, we manually detected the inner and outer boundaries of a signal zone for every 10 slices; however, for the other image slices, we determined their signal boundaries by applying an interpolation based on the boundary information detected from the selected image slices. The interpolation was effective because the boundary of the imaged colorectum shows a continuous variation along the pullback direction. Figure 4(e) presents the entire upper and lower signal boundary maps produced by the interpolation for 4000 B-scan slices. Based on the boundary information, we replaced the original signal value of a pixel $(i, j, k)$ with the lower signal value of its two adjacent left and right pixels—that is, $(i, j–1, k)$ and $(i, j+1, k)$, [see Figs. 4(b) and 4(d)], only for those pixels that are located outside the signal zone and whose signal values exceed a threshold, which was set to be approximately two times greater than the thermal noise level. The noise detection criterion was reasonable because there was no meaningful signal outside the segmented signal zones [Fig. 4(e)], and it was rare for all the consecutive three pixels along the transverse direction, i.e., $(i, j–1, k)$, $(i, j, k)$, and $(i, j+1, k)$, to all be affected by interference noise. With the algorithm, we were able to heal interference noise-affected pixels, while maintaining the delicate microvascular image information included in the signal zones that typically measured less than 1 mm in depth. After the denoising, we performed a coordinate transformation if that was necessary.

In case of the co-registered US images, we applied a different algorithm because the reverberation signals generated by the imaging window (i.e., Pebax tubing) were much stronger than the true signals generated by the intestine wall with a thickness of ~0.5 mm; thus, it was necessary to eliminate these reverberation signals to make the imaged intestine more clearly visible.

For this, we first averaged all the 4000 raw US B-scan image slices saved in a bipolar signal to extract the reverberation signals that are commonly included throughout the B-scan slices [see Fig. 4(f)]. In the averaging process, it was assumed that true signals representing the echogenicity of tissues could cancel each other out because of the randomness of the positions that generated the US echo signals. We then subtracted the common reverberation signals from all the raw US B-scan image slices to reduce its signal magnitudes in a processed US image. Of course, the subtraction process did not function perfectly because the reverberation pattern of the imaging window was not constant (note that the rotation axis of the scanning tip fluctuated rather than be stably fixed at a position during the C-scan because there was a large gap between the OD of the scanning tip and ID of the imaging window).

After the elimination of the reverberation signal, we applied the Hilbert-transform to extract the envelope of the bipolar signal and eliminated interference noises according to a similar procedure applied to a PA image. However, unlike the denoising process of a PA image, we applied a different algorithm in which a noise-affected pixel $(i, j, k)$ was detected by calculating its transverse signal gradient to its neighboring two pixels—that is, (i, j–1, k) and (i, j+1, k)—and healed it by replacing its pixel value with the lower pixel value of the two adjacent pixels. In other words, only when the two gradient values were all greater than a threshold value, we regarded it as a noise-affected pixel and the noise detection algorithm was applied throughout the entire image area of a US B-scan image. Of course, the applied noise detection criterion and healing method were reasonable because the spatial resolutions of the US mode were much poorer than those of the PA mode; thus, the signal amplitudes of imaged features in US images varied more gradually and continuously, whereas noise-affected pixels created sharp amplitude gradients along the transverse direction (note that interference noise appeared in an image as if it scratched the image vertically–that is, like raining [see the Hilbert transformed images included in Figs. 4(d) and 4(f)]).

After the denoising, we shrunk the US images along the radial direction to match the radial range with that of the PA images. Thereafter, we performed a coordinate transformation if necessary or further proceeded to an image segmentation process; the latter process enabled the further elimination of the corresponding imaging window's reverberation signals to be used for the production of a radial-maximum amplitude projection (RMAP) image or a volume-rendered image in which a better visibility of the imaged tissues was important.

Figure 5 presents the imaging results of the rat colorectum processed by the method explained above. Figure 5(a) presents a volume-rendered pseudo color image in which PA (red) and US (green) images are plotted in a merged manner (the rotating movie and a walk-through movie are available in Visualizations 1 & 2), and Figures 5(b) and 5(c) depict the US- and PA-RMAP images, respectively, processed from the same data set. Figure 5(d) is a magnified RMAP image of the dashed box area included in Fig. 5(c); Figures 5(e)–5(g) are respectively an US, merged PA and US, PA cross-sectional image taken at the dashed vertical line in Fig. 5(c), which corresponds to a probe insertion depth of ~6 cm from the anus; and Figure 5(i) is a histology image (H&E staining) of the tissue harvested around the mid-colorectum.

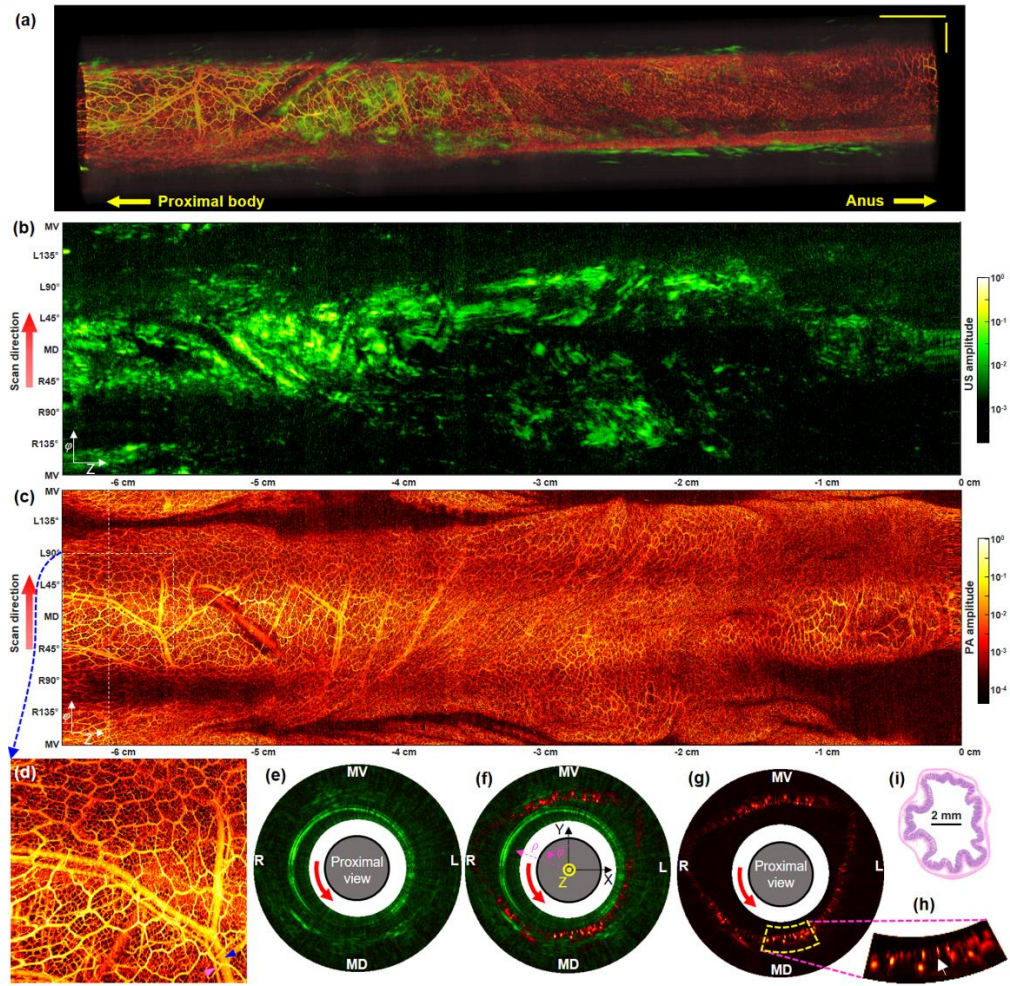

Fig. 5. In vivo rat colorectum imaging results. (a) Three-dimensionally rendered, merged PAE(red)-EUS(green) pseudo color image acquired from a rat colorectum in vivo over a ~6.4 cm range with a ~5.3 mm image diameter (Visualization 1). The right-hand side of the image corresponds to the anus. The horizontal and vertical scale bars represent 5 mm and 1 mm, respectively. (b, c) US- and PA-RMAP images of (a). The vertical axis corresponds to the angular FOV, covering 360°, and the horizontal axis corresponds to the pullback length of ~6.4 cm. MD, mid-dorsal; MV, mid-ventral; L, left; R, right. (d) Magnified view of the dashed box region in (c). Note that, in the case of (b–d), PA and US signals are mapped on a logarithmic scale. (e–g) Representative US, merged PA-US, and PA B-scan images selected from the vertical dashed line depicted in (c). (h) Magnified view of the dashed area presented in (g). (i) A typical histology image (H&E stain) of the colon.

As revealed in the two RMAP images, the two imaging modes provided different structural information according to their own contrast characteristics. As expected, the US-RMAP image provided the echogenicity distribution for the imaged tissue volume, whose signals mostly originated from the neighboring tissues that surround the colorectum. However, unlike the US-RMAP image, the PA-RMAP image revealed the vascular map of the colorectum in an incredibly detailed manner that had never been achieved previously. As depicted in Fig. 5(d), capillary networks even with a mesh size as small as ~50 μm could be visualized by this new endoscopic device. Through a close image analysis, presented in Fig. S4 of Supplement 1, we could identify that the signals corresponding to the mesh-like pattern originated from the vascular networks distributed in the innermost wall of the colorectum and corresponded to the thorn-like features revealed in the cross-sectional PA image [Fig. 5(g)] and its local magnification image [Fig. 5(h)] (see the white arrow included in Fig. 5(h)). By analyzing the thorn-like feature in the magnified image, we were able to confirm that the PA transverse resolution of the probe was lower than at least 13 μm, even in the in vivo imaging (it must be noted that any capillary scanned along the cross-sectional direction by the probe must appear in such a thorn-like shape in an image because the radial resolution is at an AR level rather than at an OR level. Through the analysis presented in Fig. S4 of Supplement 1, it is conjectured that the two large blood vessels indicated by the pink and blue arrows [Fig. 5(d)] respectively correspond to the artery and vein that supply blood to the colorectum.

From a technical viewpoint, the visualization of such a mesh-like structure was a somewhat unexpected result because the 13-μm transverse resolution of this device was even slightly worse than that (10 μm) of the previous micro-motor-based endoscopic probe reported in Ref. 18. Moreover, there was large gap between the OD of the scanning tip and the ID of the imaging window [Fig. 5(e)], which could cause an unstable fluctuation of the rotation axis unlike the micromotor-based built-in scanning mechanism. We attribute the presented capillary network visualization achievement solely to the increased frame speed of the current system, which was 5 times faster than that of Ref. 18, thereby the connectivity of microvascular networks in adjacent image slices could be secured. However, regardless of the technical basis, considering the transverse resolution (~13 μm) of the probe, typical sizes of RBCs (~7–8 μm), and the apparent hole size of the mesh structure with a fairly uniform hole size (~50 μm), it is presumed that the visualized structure is the "lowest level" capillary network distributed in the colorectum and that a kind of morphological hierarchism exists in the mesh-like capillary networks developed in the colorectum, which refers to the theory presented in Ref. 42 (see the related rationale presented in Fig. S4 of Supplement 1).

In presenting the US-and PA-RMAP images, we applied a logarithmic color scale, through which approximate SNR also could be estimated, to better present imaged structures that revealed a large signal variation caused by the large difference in the echogenicity [Fig. 5(b)] and the target distance with respect to the fixed optical WD [Fig. 5(c)]. Although there was such a target distance variation, it must be noted that the high-resolution and high contrast visualization of the capillary networks presented in Fig. 5(d) was possible because the imaged tissue area was fairly positioned around the optical WD. In Fig. S5 of Supplement 1, we present our analysis result showing the PA signal variation according to target distance, whose trend fairly agreed with the optical WD measurement result that was performed using a blade as depicted in Fig. S1 of Supplement 1. In the case of presenting the US [Fig. 5(e)] or merged PA and US B-scan images [Fig. 5(f)], the profile of the imaging window was not excluded, although their magnitudes were still much greater than those generated by intestinal tissues even after the membrane reverberation signal subtraction process [Fig. 4(f)]. However, in producing the US-RMAP image [Fig. 5(b)], we applied the imaging window signal elimination procedure illustrated in Fig. 4(f) to better display the echogenic feature of neighboring tissues that surrounded the colorectum. A pullback movie revealing similar cross-sectional images over the first 1,000 slices are presented in Visualization 3.

In the experiment, we induced and maintained the anesthesia of the animal by using isoflurane (4% for induction, 1.5%–2.0% for maintenance), injected medical US gel into the colorectum for acoustic matching before every probe introduction, euthanized the animal by using $CO_2$ gas after acquiring five C-scan data sets, and finally harvested the imaged organ for validation. Before endoscopic imaging, the animal was fasted for 24 hours to reduce the amount of feces in the colorectum. All procedures in the experiment followed protocols approved by the Institutional Animal Care and Use Committee at UNIST (UNISTIACUC-20-53).

**C. Swine esophagus imaging in vivo via the instrument channel**

Since the major goal of this study was to develop an integrated PAE-EUS mini-probe system that can be promptly applied, just like the conventional commercialized EUS mini-probes [9,10], when a suspicious tissue is found in a GI tract, we performed an experiment to image the upper GI tract of a swine (Yorkshire, 3 months, 35 kg, OrientBio, Seongnam, S. Korea) to demonstrate the intra-instrument channel workability of the mini-probe system.

As shown in Visualization 4, we were able to successfully acquire a series of PAE and EUS images from the esophagogastric junction of the animal in conjunction with a commercialized video endoscope (GIF-2T240, gastroscope, Olympus) within a mere 20 seconds from the initial introduction moment of the mini-probe into the entrance port of the instrument channel. Figures 6(a) and 6(b) are the still cuts of Visualization 4, taken just before the introduction of the probe to the instrument channel and just after the probe's projection from the output port. However, before introducing the mini-probe, it was important to ensure there were no air bubbles around the scanning tip by imaging a surgical blade (Visualization 5).

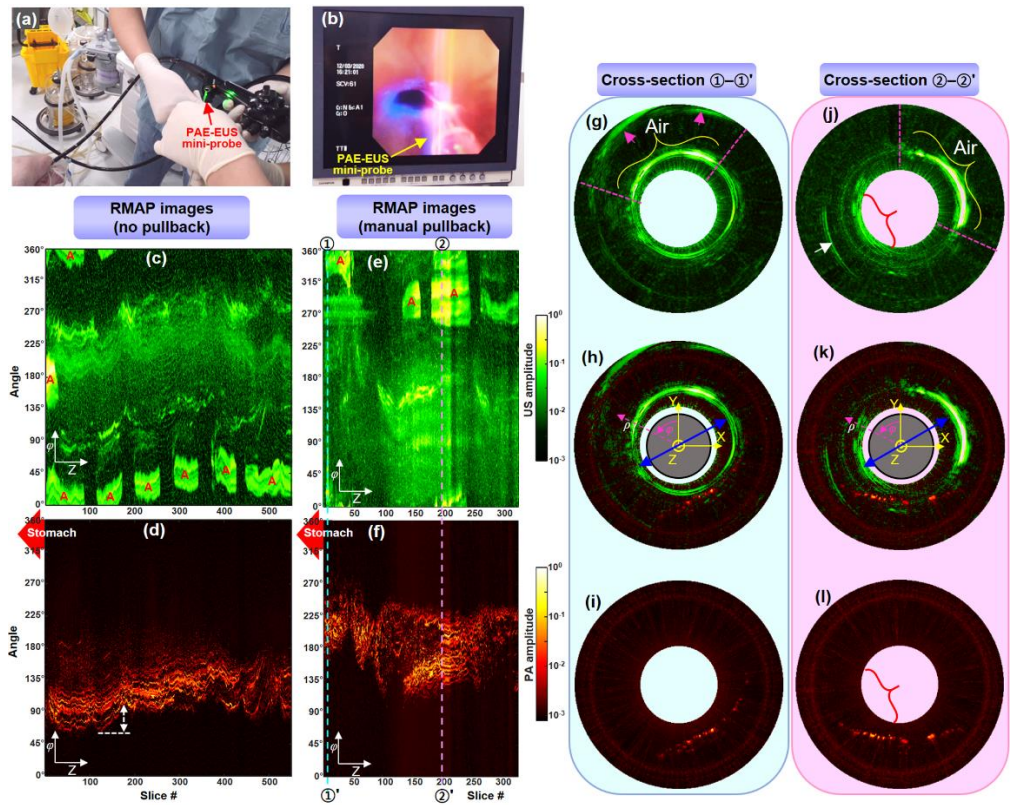

Fig. 6. Swine esophagus in vivo imaging results. (a, b) Still cuts of Visualization 4, captured just before the probe introduction to the instrument channel (a) and just after the probe's projection from the output port (b). (c, d) US- and PA-RMAP images acquired from the esophagogastric junction of the animal without any manipulation (no pullback) applied to the gastroscope. (e, f) US- and PA-RMAP images acquired from another spot of the esophagogastric junction by manually pulling back the gastroscope in which the mini-probe was introduced. In (c) and (e), "A" represents the local surface region of the Pebax imaging window that interfaced with air. (g–l) Two sets of US, merged PA and US, and PA B-scan images selected from the two vertical dashed lines depicted in (e) and (f) along ①–①' (g–i) and ②–②' (j–l). These images are views at the proximal of the probe.

We were able to acquire over 20 data sets from different sites from the animal, such as mid- and lower-esophagus and the greater curvature of the stomach. However, the desired image acquisitions were not always favorable because they were mostly subjected to the associated acoustic coupling condition that solely responded to the body liquid and intermittent waterjet; moreover, there were strong motion artifacts. These two were truly the main aspects that affected the quality of the acquired images. Moreover, due to the tightness of the mini-probe within the instrument channel, we were unable to perform a motorized pullback C-scan, as we did with the rat colorectum imaging. Among the acquired data sets, here we present results processed from two data sets—one from the data set acquired continuously at the same position but with no scope manipulation and another data set that was acquired by manually pulling back the entire scope set, in which useful PA signals acquired from the esophagogastric junction were included.

Figures 6(c) and 6(d) depict the US- and PA-RMAP images, respectively, processed using 550 B-scan image slices that were acquired from the first procedure. As indicated in the PA-RMAP image [Fig. 6(d)], it is evident that the blood vessels were not mapped straight, although the operator did not manipulate the gastroscope in any manner. It was because of the motion artifacts that were mostly caused by the breathing motion. The patch-like regions denoted by the letter 'A' in the US-RMAP image [Fig. 6(c)], which represent hyperechoic signal regions generated by a polymer-air interface of the imaging window, manifest the breathing cycle of the animal in intervals of ~4–5 seconds. As indicated by the white arrow presented in the PA-RMAP image [Fig. 6(d)], the transverse image distortion was as great as 1.3 mm for one breathing cycle.

Even though we attempted the manual pullback for the entire scope set, we were unable to acquire a reliable vasculature map, as presented in Fig. 6(f), which is a PA-RMAP image processed from 320 B-scan slices. By analyzing the PA B-scan images utilized for the PA-RMAP image [see Visualization 6], we found that the transverse displacement of a blood vessel appeared in two successive images (i.e., in intervals of 50 ms) and was occasionally even greater than 200 μm, which far exceeded the blood vessel diameter that can be resolved by the current system. In the corresponding US-RMAP image [Fig. 6(e)], strong acoustic reflections also appeared at multiple sites because the acoustic coupling was not stable during the manual pullback, which appeared to be attributed to the uneven surface condition of the imaged tissues under the unstable contact condition of the small diameter mini-probe to the esophageal wall, which had a larger diameter [Fig. 6(b)].

A close image analysis revealed that hyperechoic responses of the imaging window appeared in approximately two cases; 1) when the corresponding angular portion of the imaging window did not make contact with the tissue but with air or 2) when the membrane surface of the imaging window was perpendicularly oriented to the incident acoustic beam axis. Whether a hyperechoic signal was generated by ambient air or by the vertical incidence of an acoustic beam to the membrane surface could be judged by comparing co-registered PA-RMAP image and also by browsing US B-scan images slice-by-slice. Whichever the case, most often, a corresponding reverberation artifact appeared once more in the US B-scan image at the position corresponding to one round trip travel time of US waves between the US transducers and the imaging window. However, in the first case, no PA signal was observed at the corresponding positions in the PA-RMAP image. Moreover, there was a tendency that such hyperechoic signals were generated regardless of the angle between the US beam axis and the imaging window, and the corresponding echogenicity was much greater than that of the latter case.

The two set of US, merged PA and US and PA B-scan images [Figs. 6(g)–6(l)] taken from the two dashed lines depicted in Figs. 6(e) and 6(f), represent related examples for the first case. As depicted in the first US B-scan image [Fig. 6(g)] taken at ①–①', a hyperechoic response of the imaging window occurred on account of the polymer-air interface present between the 10 o'clock and 1 o'clock positions, although the acoustic beam axis was not perpendicular to the membrane surface and its reverberation artifact also appeared at a farther position radial-wise (see the feature indicated by the pink arrow). The blue arrow assigned in the merged PA and US image [Fig. 6(h)] indicates the direction along which the rotation axis of the scanning tip was deviated from the center of the imaging window (see its boundary visualized in the image). In the second US B-scan image, [Fig. 6(j)] taken at ②–②', another type of hyperechoic response of the imaging window caused by polymer-air interface occurred between the 12 o'clock and 4 o'clock positions. In other words, in this case, not only a polymer-air interface was involved but also the membrane was perpendicular to the acoustic axis, as indicated by the blue arrow in Fig. 6(k). Whichever the case, the fact that it was surely caused by a polymer-air interface could be confirmed by "no signal presence" at the corresponding positions in the PA images [Figs. 6(i) and 6(l)].

Furthermore, a hyperechoic signal generation caused only by the vertical incidence of the acoustic beam to the membrane revealed a different pattern. As indicated by the red marker located in the 8 o'clock direction in Fig. 6(j), the acoustic response of the corresponding membrane portion was lower in amplitude than that of the polymer-air interface and its perpendicularity to the rotating acoustic beam axis could be assured from the orientation of the blue arrow in Fig. 6(k). Although the amplitude was lower, a corresponding reverberation artifact still appeared at a farther distance (see the feature indicated by the white arrow). In this case, since there was no acoustic blockage caused by a polymer-air interface and useful signals could still be acquired through the region colored in red in both the US [Fig. 6(j)] and PA [Fig. 6(l)] images. Unlike the US B-scan images, in producing the US-RMAP images [Figs. 6(c) and 6(e)], we eliminated the signals corresponding to such a reverberation artifact (i.e., signals indicated by the white arrow in Fig. 6(j)) and the imaging window as well by manually segmenting their locations in each B-scan image to better present other image features that appeared weakly.

Unlike the image processing of the rat colorectum, it was difficult to satisfactorily eliminate the reverberation signals generated by the imaging window by applying the simple subtraction method illustrated in Fig. 4(f); this is because the number of available US B-scan slices was much lesser and the reverberation pattern was not constant as the rotation axis of the scanning tip highly fluctuated, even seriously deviating from the center of the imaging window. Moreover, as depicted in Fig. 6(j), a certain portion of the imaging window was not even completely visualized because it was not within the stipulated window of the gated time, as the operator increased the angulation of the gastroscope's distal end to achieve a better acoustic coupling between the mini-probe and the target tissue during the procedure. In terms of imaged features in the acquired PA B-scan images [Figs. 6(i) and 6(l)], thorn-like shapes as were seen in the rat colorectum image were barely observed in the epithelium of the imaged esophagogastric junction. Although the target distance was highly fluctuating during the procedure, we attribute it to the intrinsic feature of vascular pattern and diameter depending on the type of organ [43]. A movie showing the 320 B-scan slices acquired through the manual pullback is presented in Visualization 6.

Apart from the presented results, we were able to acquire additional data from other sites, as mentioned earlier, until the engaged probe sheath was damaged—there was a water leakage through the UV glued section of the two tubings [see Fig. 1(b)], particularly when we approached the stomach region, which required a higher angulation of the gastroscope. However, based only on acquired B-scan image information, it was difficult to find a notable difference in terms of morphological features for blood vessels that were acquired from different anatomical sites—for example, the esophagus and stomach.

In the experiment, we anesthetized the animal by injecting (IV) a drug cocktail comprising Alfaxalone (5 mg/kg), Azaperone (5mg/kg), and Xylazine (1mg/kg); maintained the anesthesia by using isoflurane (2% in 2 L/min, 50% oxygen); and subsequently euthanized the animal with a KCL (2 mmol/kg) injection (IV). Furthermore, prior to endoscopic imaging, the animal was fasted for 24 h to reduce the amount of ingesta in the stomach. All procedures in the experiment followed protocols approved by the Institutional Animal Care and Use Committee at Korea University (KOREA-2017-0128-C1).

## 3. Summary and Discussion

In this article, we presented the first torque coil-based OR-PAE and EUS mini-probe images acquired from a swine esophagus in vivo along with a rough scheme for related image processing and interpretation, which could be informative for follow-up studies on GI endoscopy application. Although the implemented system needs to be further improved in terms of hardware and software, this study showcased the potential benefit of a complementary image production attributed to the PAE and EUS techniques. Moreover, through the rat colorectum imaging study, we showcased an approximate level of 3D vascular visibility which would be attainable by PAE in an actual clinical practice in the future. To the best of our knowledge, the presented vascular map was the first one that was acquired from virtually the entire area (longitudinal and angular) of the colorectum of a live rat based on the highest PAE-resolving power ever achieved and thereby enabled in the in vivo visualization of the arteriolar and venular networks that suggested their morphological hierarchism [42] expressed in the colorectum [Fig. S4 of Supplement 1] (note that conventional capillary-level vascular morphology investigation studies, of course for the colorectum as well, mostly relied on invasive approaches, like vascular corrosion casting [44]).

Early detection of angiogenesis, a phenomenon in which malignant tissues continuously secrete substances that induce neovascularization around them to receive nutrients and oxygen needed for rapid growth, is an important issue in real-world practice. As more than 85% of all cancers originate in the epithelium that lines the inner surfaces of organs throughout the body [45], and also since such a vasculature pattern is regarded as an important indicator for pre-malignant or malignant lesion, high-resolution mapping of a vasculature distributed in the epithelium of the GI tract would be diagnostic for early cancer detection as well as for related clinical decisions. Apart from the clinical perspective, the presented endomicroscopic technique could inspire animal model-based colorectal tumor microenvironment studies [46,47] on account of the label-free visualization of capillaries in vivo and in situ.

The high-resolution visualization of capillary networks as low as ~13 μm was possible due to the satisfactory operation of the optical rotary junction and the distal optics implemented based on a 10-μm core diameter optical fiber, which was the narrowest among all the previous in vivo-applicable, torque coil-based PAE-EUS probes or IVPA catheters reported thus far [19,20,28–30,32–34]. Most of all, as the deposited laser pulse was more tightly focused inside a target tissue, the required laser pulse energy could also be reduced to less than 1 μJ, which enabled a laser dose reduction effect of more than one order of magnitude compared to those applied to the previous IVPA probes that were implemented based on larger core diameter optical fibers (typically >100 μm core) [28–30,32–34]. We believe that this was also possible because the illumination optics and the dual transducers were adequately assembled within an acceptable tolerance. Moreover, the hermetic sealing of the acoustic matching medium as well as the applicability of the rotary transformer-based signal coupling mechanism near the proximal has been successfully demonstrated for the integrated PAE-EUS imaging technique, which was also the first of its kind.

However, along with its positive aspects, this study also showed a fundamental limitation—that a continuous 3D vasculature image could not be acquired from the in vivo swine esophagus imaging experiment. Considering that the key benefit of PAT is that it is capable of visualizing 3D vasculature, the presentation of simple B-scan images would not be so useful for an endoscopist. Through the first intra-instrument channel-based swine esophagus imaging experience, we recognized that a simple improvement in the scanning speed (e.g., >30 Hz) would not be sufficient to acquire the continuous 3D vasculature map because a stable 3D data acquisition, if it is still based on the manual pullback of the entire sheath of the mini-probe through the instrument channel, would not be fundamentally possible in an actual clinical environment in which strong motion artifacts are frequently involved. In this regard, the development of a next-generation endoscopic system with a rapid

self-3D internal helical scanning mechanism just inside an engaged probe sheath would be most essential.

Furthermore, in undertaking this study, we did not hesitate to diminish the voltage level of the US excitation pulse because we believed that, in terms of the dual-modality-based image interpretation for a target tissue, it would not be necessary to acquire US images over a deeper region than the co-registered OR-PAE image, which has a very shallow image depth—typically less than ~1 mm. However, through this study, we recognized that it was a better direction to harness the deep imaging capability of US imaging for the purpose of visualizing neighboring anatomical landscape over a radial range that is as wide as possible. In this case, the signal balancing issue could possibly be resolved by applying a different gain to each imaging mode while maintaining the conventional excitation voltage level (>100V) for the US imaging mode, rather than merely sacrificing the excitation voltage (as done in the current study). In addition, finding a method to minimize interference noise, centralize the rotation axis of the scanning tip in a stable manner, and embody a dedicated probe sheath with different wall thicknesses between the imaging window and the body section (which also includes braided metallic wires) would accelerate the clinical translation of the integrated PAE-EUS technique.

**Funding.** U-K Brand Research Fund of UNIST (1.210039.01); Korea Medical Device Development Fund grant funded by MSIT, MOTIE, MOHW, and MFDS (1711138075, KMDF_PR_20200901_0066); National Research Foundation of Korea (2015R1D1A1A01059361, NRF-2019R1A2C1004274, NRF-2020R1A2C4002621).

**Acknowledgments.** The authors thank Y.J. Choi for the assistance with histology and UCRF (UNIST Central Research Facilities) for support of using the equipment. This study contains the results obtained by using the equipment of UNIST Central Research Facilities (UCRF).

**Disclosures.** The authors declare no conflicts of interest.

**Data availability.** The data that support the findings of this study are available from the corresponding author upon reasonable request.

**Supplemental document.** See Supplement 1 for supporting content.

# Intra-instrument channel workable, optical-resolution photoacoustic and ultrasonic mini-probe system for clinical gastrointestinal endoscopy: supplemental document

## 1. Measurement of the optical working distance (WD) and PA and US spatial resolutions

To find the WD of the OR-PAE mode and to measure the PA and US spatial resolutions, we imaged a surgical blade (#4) submerged in a water tank according to the setup shown in Fig. S1(a). Figures S1(b) and S1(c) are schematics showing the scanning configuration viewed at the proximal of the mini-probe and the structure of the scanning tip, respectively.

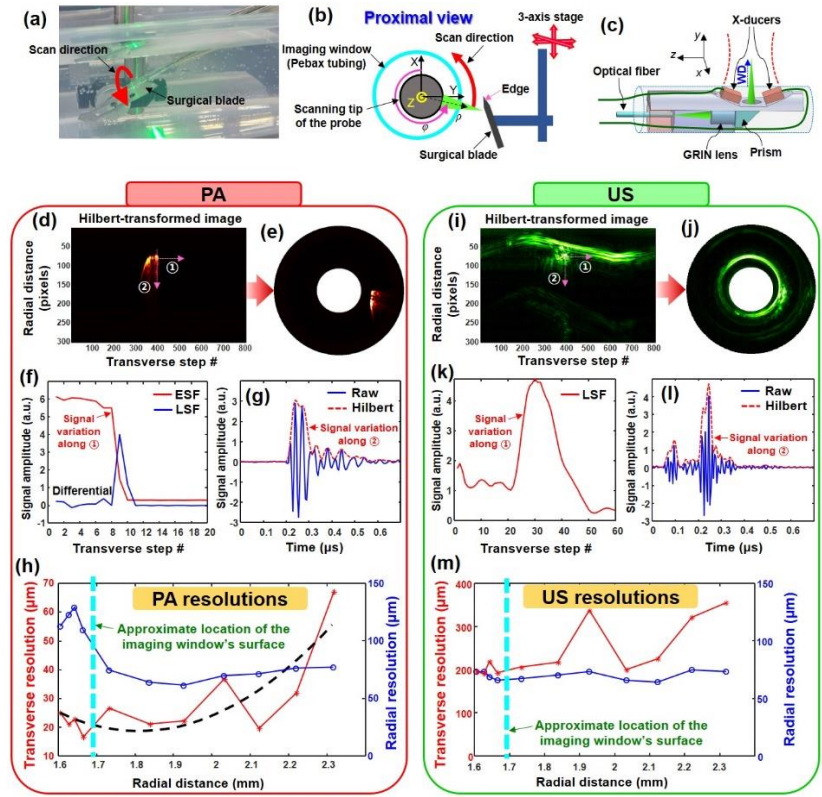

Fig. S1. Measurement of the optical WD and PA and US spatial resolutions. (a) Photo showing the experimental setup. (b, c) Schematics showing the scanning configuration viewed at the proximal of the mini-probe (b) and the structure of the scanning tip (c). (d, e) Hilbert-transformed PA B-scan images (averaged by 100 B-scan slices) plotted in rectangular (d) and circular format (e). (f) PA ESF and LSF extracted from the blade image along direction ① shown in (d). (g) Raw and Hilbert-transformed PA A-line signals generated by the blade along direction ② shown in (d). (h) PA resolution variation graph according to the target distance. (i, j) Hilbert-transformed US B-scan images (averaged by 100 B-scan slices) plotted in rectangular (i) and circular format (j). (k) US LSF extracted from the blade image along direction ① shown in (i). (l) Raw and Hilbert-transformed US A-line signals generated by the blade along direction ② shown in (i). (m) US resolution variation graph according to target distance.

In general, it is accepted that the transverse resolution of an OR-PAT device appears much better than the radial resolution because it is mainly determined by the laser beam diameter at a target distance. However, in the case of the radial resolution, it usually gives a value of several tens micrometer because it is not determined by the optical parameters but by the acoustic parameters, such as center frequency, bandwidth, and acoustic NA, as with the cases of the radial and transverse resolutions of the US imaging mode. Thus, if the sharp edge of the blade is imaged at different positions, it is possible to know the resolution variations according to the target distance by analyzing the acquired images along the transverse and radial directions.

For the experiment, we affixed the blade onto an aluminum holder [Fig. S1(b)], whose position could be precisely controlled by a 3-axis stage, with a proper tilting angle, as shown in Fig. S1(b), so that the surface of the sharp edge could be perpendicularly oriented to the optical beam axis, and then we acquired images at 11 different positions. However, to minimize measurement errors caused by the laser fluctuation, we recorded 100 B-scan images at each position.

Figs. S1(d)–S1(g) show a set of experimental results that was selected to illustrate the related data analysis procedure for the PA imaging mode and that corresponded to the $4^{th}$ target position, and Fig. S1(h) represents a PA resolution variation graph acquired according to the explained procedure for all the target positions (Figs. S1(i)–S1(l) and Fig. S1(m) are corresponding US counterparts, respectively).

In brief, we first averaged all the 100 B-scan images acquired at a position for each imaging mode and performed a Hilbert transform [Figs. S1(d) and S1(i)] to extract the envelope of the bipolar signal (Figs. S1(e) and S1(j) are corresponding coordinate-transformed images). We then extracted signal values along the transverse (①)/radial (②) directions, as depicted in Figs. S1(f) and S1(k)/Figs. S1(g) and S1(l) are corresponding signal variation graphs. Here, it is a natural interpretation that the extracted signal variation curves (red) around the sharp edge of the blade correspond to an edge spread function (ESF) for the PA imaging mode [see the red curve in Fig. S1(f)] and a line spread function (LSF) for the US imaging mode [see the red curve in Fig. S1(k)], respectively, because the relative sizes of the blade edge to the optical and acoustic beam diameters are different for the two imaging modes. Although an ESF was acquired for the PA mode [Fig. S1(f)], we could obtain an LSF by taking its differential, from which a transverse resolution has been estimated based on the FWHM.

Figures S1(h) and S1(m) represent the PA and US resolutions, respectively, determined by the explained procedure for the 11 target positions. As appeared in the PA resolution curves [Fig. S1(h)], the transverse resolution variation according to the target distance curve (red) did not show a smooth behavior as with the radial resolution (blue), and all the measured resolution values appeared much greater than the optical beam diameter (~8 μm) determined using the optical beam profiler: Its minimum value appeared to be 17 μm at the $4^{th}$ position. We attribute the large fluctuation of the transverse resolutions to the unideal measurement condition caused by the relatively larger transverse step size of the scanning tip (~10–30 μm; note that this value varies depending on a target distance) compared to the optical beam diameter (~8 μm) and the instability (i.e., jitter) of the rotation axis of the scanning tip during the 100-image slice acquisition. Thus, to find a reasonable optical WD based on a curve fitting, we employed a second order polynomial (black dashed line) and determined the WD to be ~1.8 mm from the center of the scanning tip. This means that, with respect to the surface of the imaging window, the optical WD was located only 0.1 mm apart from it (see the vertical dashed line added in the graph to denote the location of the imaging window), which means that the laser beam's focal spot traced a circle with a ~3.6-mm diameter.

Although the transverse resolutions determined based on this method were all greater than the value expected based on the optical beam diameter at the focus [Fig. 1(e)], we claim that the actual resolution of the PA imaging mode was as low as ~13 μm because we could resolve capillaries with diameters comparable to the claimed value, as presented in Fig. S4.

In the case of the radial PA resolutions [see the blue curve Fig. S1(h)], they demonstrated a relatively smoothly varying behavior around the ~70 μm value over the zone outside the imaging window; however, they displayed an increase inside the imaging window as the radial distance decreased. We attribute the near-field radial resolution degradation to the acoustic geometric aberration caused by the employment of the dual transducers.

The US resolutions displayed a gradually increasing tendency for the transverse resolutions ranging from 191 to 355 μm, although with some fluctuations [Fig. S1(m)]. However, the radial resolutions exhibited almost uniform values around the ~70-μm level over the investigated radial distance. In the current geometric configuration of the dual transducers [Fig. S1(c)], it should be noted that spatial resolutions between the transverse (x) and longitudinal (z) directions for the US mode must be different because, although weak, there was an acoustic focusing along the y-z plane but not in the x-y plane. Although we tried to measure the longitudinal resolutions, we could not acquire a reliable result because it was difficult to maintain the rotation axis of the scanning tip stable during the pullback scan.

In summary, based on the presented measurement method, transverse resolutions appeared to range from 17 to 67 μm for the PA imaging mode and from 191 to 355 μm for the US imaging mode. Radial resolutions ranged from 61 to 130 μm for the PA imaging mode and from 66 to 75 μm for the US imaging mode over the investigated radial distance.

## 2. Measurement of transmission efficiency of the rotary transformer

We measured the transmission efficiency (or insertion loss) of the rotary transformer according to the distance of physical separation between the rotor and stator using the procedure illustrated in Fig. S2. This experiment was also intended to investigate how much longitudinal runout could be allowed for the ball bearing installed in the electrical rotary joint [Fig. 2(e)]. Since the turns ratio of the first and second coils was 1, we employed a 50-Ω resistor as a load to be engaged to the 50-Ω output port of a function generator (FG, model #: AFG2021, Tektronix, USA).

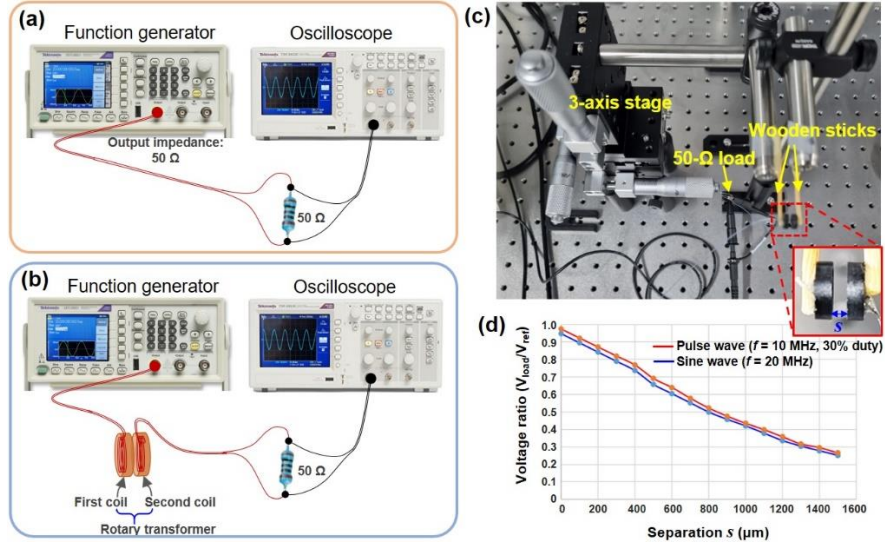

Fig. S2. Rotary transformer transmission efficiency measurement. (a) Setup to set a reference waveform. (b) Setup to measure the amplitude of the voltage delivered to the 50-Ω load. (c) Photo showing the experimental setup constructed to measure the transmission efficiency of the rotary transformer. (d) Experimental results.

First, to set a reference voltage amplitude ($V_{ref}$) and frequency to be applied to the 50-$\Omega$ load, we connected the load directly to the output port of FG, as shown in Fig. 2S(a) and set FG to generate a sinusoidal wave with a 200 mV$_{pp}$ and a frequency of 20 MHz, whose waveform was also verified by an oscilloscope (TDS2022C, Tektronix).

We then inserted the rotary transformer between the FG and the 50-$\Omega$ load, as shown in Fig. 2S(b), and measured the variation of the voltage ($V_{load}$) transferred to the load according to the separation ($s$) between the rotor and stator by using the oscilloscope. Figure 2S(c) shows the photo of the rotary transformer in which the stator part was fixed and the rotor part was mounted on a 3-axis stage.

As depicted in Fig. 2S(d), the rotary transformer exhibited roughly 95% efficiency for one-way signal transmission when $s = 0$, and the efficiency gradually decreased to 70% as $s$ increased to 400 μm. We also performed a similar measurement after changing the waveform to a pulse wave with a 200 mV$_{pp}$ and a frequency of 10 MHz (duty: 30%) and found that the overall decay trend was almost same with the sinusoidal waveform case; however, the overall efficiencies appeared slightly better [see Fig. 2S(d)].

### 3. Circuitry details

To provide further information on the circuitry of the implemented system, we present Fig. S3. Figure S3(a) depicts the entire block diagram, which is almost the same as Fig. 3(a), and Fig. S3(b) shows corresponding photos. In Fig. S3(b), the two insets are the magnified views of the RF switch and signal amplifier unit that was implemented in a stacked structure and the distal end of the mini-probe in which the dual transducers were affixed.

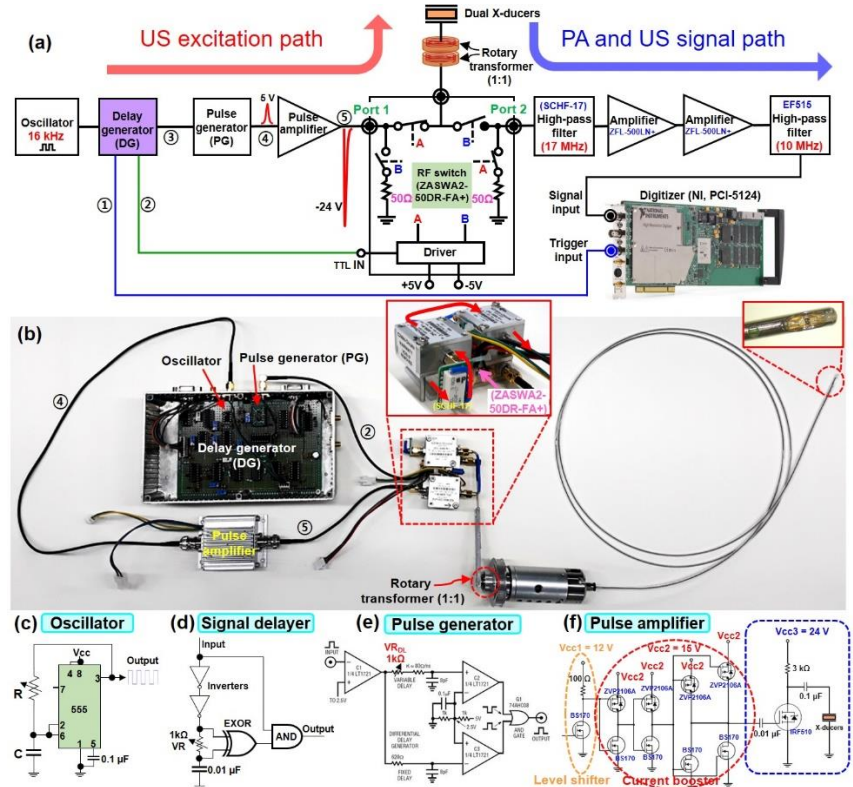

Fig. S3. Circuitry of the integrated OR-PAE and EUS mini-probe system. (a, b) Block diagram (a) and photo (b) of the implemented circuit. (c) Circuit diagram applied to the oscillator. (d) Signal delayer circuit concept applied to the delay generator. (e, f) Circuit diagrams applied to the pulse generator (e) and the pulse amplifier (f). The circuits presented in (f) were drawn based on Reference 41.

Figs. S3(c)–S3(f) are the circuit diagrams applied to the oscillator, signal delayer, pulse generator, and pulse amplifier, respectively. The oscillator was implemented based on the well-known astable oscillator circuit using a NE555 timer IC. As the oscillation frequency was determined by 0.722/RC, it was possible to embody a variable-frequency oscillator by applying a variable resistor (VR). We also embodied the delay generator (DG) shown in Figs. S3(a) and S3(b) by applying the principle of a propagation delay generated by the capacitor and VR included in Fig. S3(d). The pulse generator (PG) was implemented based on LT1721 (Linear Technology, USA), and its pulse duration could also be adjusted by adding a $VR_{DL}$. For the pulse amplifier, which played the role of a pulser in the current embodiment, it was implemented based on the high-voltage pulse amplification concept described in Reference 41 (we referred to the level shifter and current booster circuits described in the reference). In Fig. S3(f), the voltage Vcc3 applied to the final stage determined the output pulse amplitude of the pulse amplifier. Indeed, as described in the reference, the output voltage level of the pulse amplifier could be arbitrarily attained by applying a proper DC source. In the current study, however, we applied 24 V because of reasons explained in the main text. The element "X-ducers' depicted in the final stage represents the dual US transducers (i.e., load) installed in the distal end of the mini-probe. Of course, the generated 24V pulse was sent to the transducers via the RF switch (ZASWA2-50DR-FA+) when the relevant electric route was created.

## 4. Capillary network analysis

To clarify the depth-wise location of the smallest mesh-like capillary networks and to investigate the morphological variation of the imaged vasculature according to depth, we closely analyzed the data utilized for Fig. 5(d). In Fig. S4, the panel named "Whole data" is the same as Fig. 5(d).

Since the sampling rate was 200 MHz, one radial data point corresponded to 5 ns in the time domain and ~8 μm in the spatial domain when assuming the speed of sound in tissue to be 1.55 mm/μs. Thus, based on surface boundary information presented in Fig. 4(e), we were able to analyze the depth-dependent vasculature variation for the "Whole data" by peeling the recorded signals layer by layer with a 5-ns interval (or ~8 μm in depth).

First, through the analysis, we identified that the mesh-like capillary networks were distributed in the innermost surface of the colorectal wall and with a fairly uniform size, as shown in the panel Depth 1, whose data included PA signals covering ~46 μm in depth from the innermost surface of the colorectum. In addition, as shown in the panel Depth 1, a faint profile of one step-larger mesh-like structure (i.e., the vessel indicated by the white arrow) was also slightly included in the panel Depth 1 and whose profile became more obvious in the panel Depth 2. Furthermore, the appearance of the artery and vein pair, which were marked by the pink and blue arrows in panel Depth 7, also began to appear from panel Depth 2 and was included up to panel Depth 15. Here, we named the smallest mesh-like capillary networks "Level 1 mesh" and the one step-larger mesh-like structure as "Level 2 mesh."

Through observation of the transverse width of the individual capillaries visualized in the panel named Depth 1, we found that they only occupied a single pixel, that is, one angular step size of the scanning tip. Considering the approximate radial distance (~1.7 mm) of the imaged area from the center of the scanning tip, one pixel corresponded to a ~13 μm displacement along the transverse direction (note that the imaged tissue area directly contacted the surface of the imaging window with a 3.38-mm outer diameter). Through this calculation, we could be convinced that the transverse resolution of the device was at least lower (i.e., better) than 13 μm, even for the in vivo imaging. Of course, in the current study, we were unable to fully harness the full PA resolving capability of the device, which is usually determined by the optical beam diameter (8 μm) [Fig. 1(e)], because we operated the system at 800 steps/revolution to attain a 20-Hz B-scan frame rate under the limited PRR (16 kHz) of the employed laser system.

In relation to the diameters of the individual capillaries, it should also be noted that PA signals corresponding to the mesh-like capillary networks began to appear from panel Depth 1, whose data corresponded to a depth of ~46 μm, to panel Depth 8, which corresponded to a depth of ~147 μm. When considering the capillary diameter estimated above (i.e., 13 μm), this result may seem slightly unreasonable. However, this result can be explained by the presence of a depth-wise elongation of the capillary PA signals caused by the relatively poor radial resolving capability of the PA imaging mode compared to the transverse resolution (note that the radial resolution was determined to be ~70–80 μm in Fig. S1).

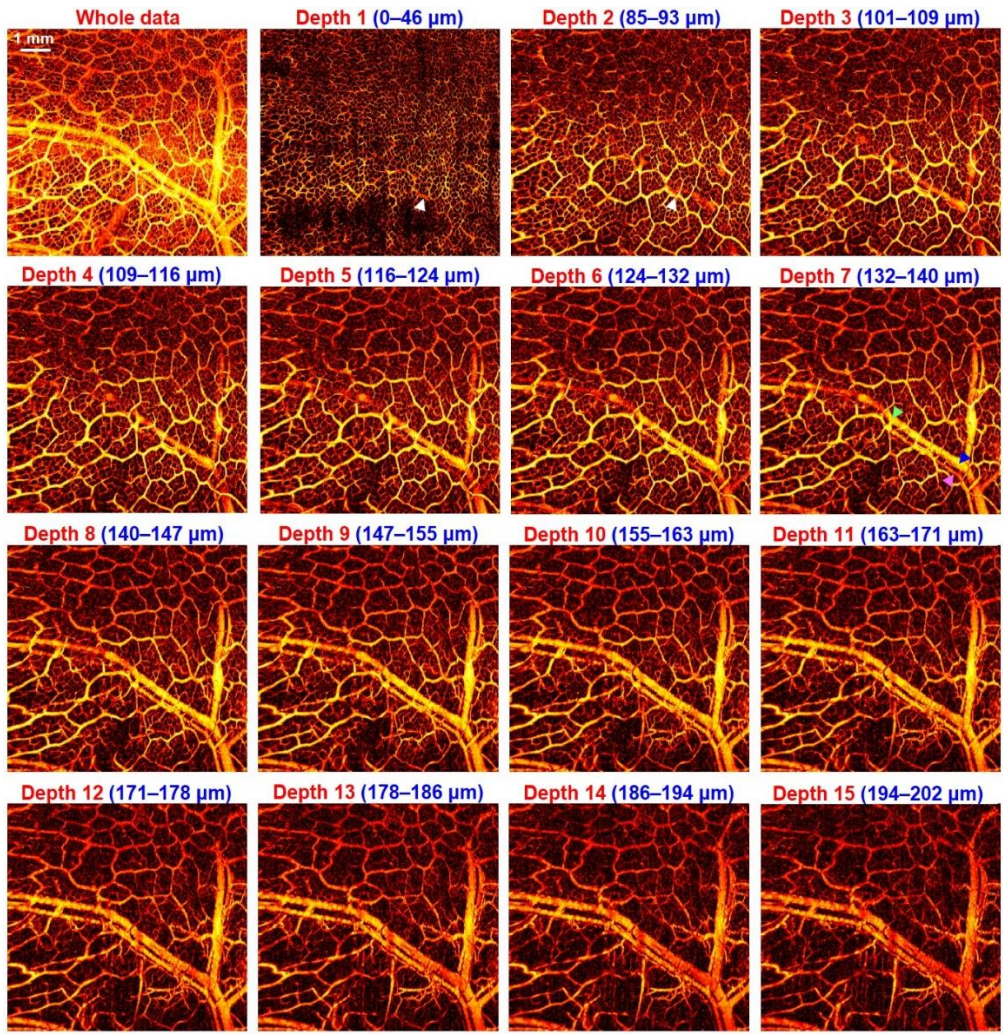

Fig. S4. Capillary network analysis.

Another important observation that needs to be mentioned based on the images presented in Fig. S4 is that there appeared to be morphological hierarchism in relation to the mesh-like vascular networks developed in the colorectum. A close observation of the morphological variation of the vasculature revealed that there appeared to be approximately two leveled mesh-like vascular networks, referred to as Level 1 mesh and Level 2 mesh in this article in accordance with their mesh hole size. We also found that Level 1 mesh with a hole size of ~50–70 μm is a sub-branched network of the artery vessel (see the pink arrow) shown in the panel labeled Depth 7, and Level 2 mesh with a hole size of approximately ~1 mm is a sub-branched network of the vein vessel (see the blue arrow). This interpretation was encouraged by the observation that the green arrow in panel Depth 7 appeared to be a bifurcation point between the vein (i.e., the vessel marked by the blue arrow) and the Level 2 mesh as well as by similar observations reported in previous studies that such an artery (pink arrow) and vein (blue arrow) pair usually run parallel to each other and that an artery vessel typically appears narrower than a vein [1]. Based on this observation, we plan to conduct a follow-up study.

## 5. Relationship between target distance and PA signal amplitude

Although the WD of the OR-PAE mode was already estimated in Fig. S1, it was possible to further verify the WD by analyzing the relationship between the PA signal amplitudes and the colorectal wall distances for the in vivo rat colorectum imaging data as presented in Fig. S5, because the colorectal wall distance was not constant but gradually varied with respect to the optical WD of the PA imaging mode. Thus, by combining the two types of information, we could further confirm the optical WD, which was determined in Fig. S1, based on the blading imaging method.

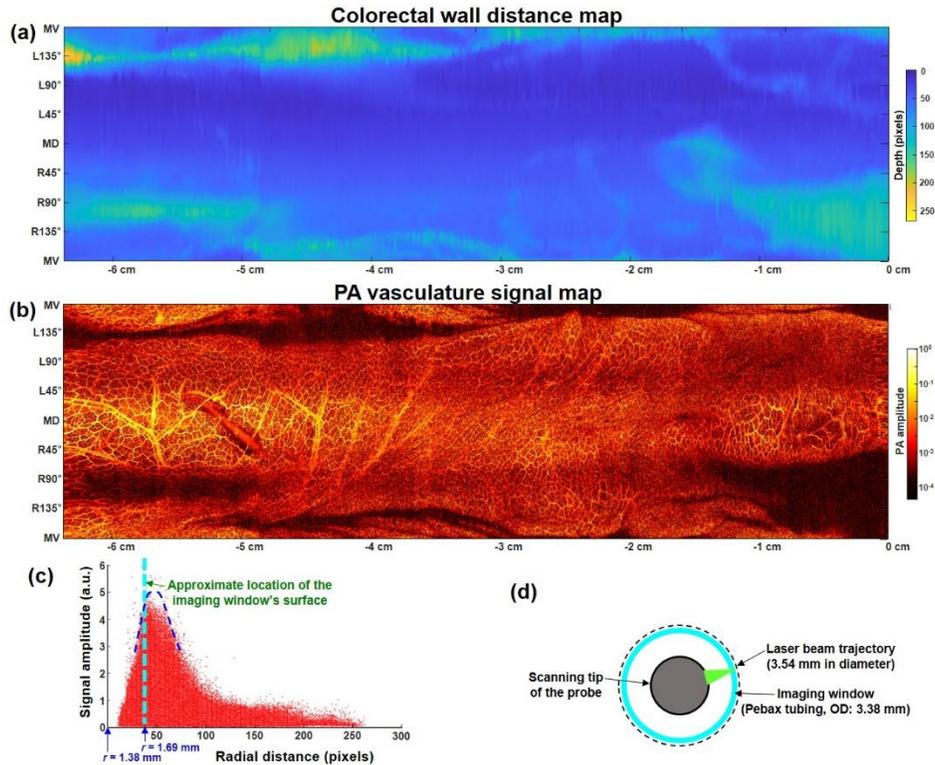

Fig. S5. Analysis of the signal amplitude variation of the OR-PA image mode in accordance with the target distance. (a) Colorectal wall distance map extracted from the PA image data set (b) acquired from a rat colorectum. (c) PA signal amplitude vs. radial distance. (d) Illustration of an assumed laser beam trajectory.

Figures S5(a) and S5(b) are the same images presented in Figs. 4(e) and 5(c), and they respectively correspond to a colorectal wall distance map and a PA vasculature signal map. However, as we combined the two types of information, a signal variation according to a target distance appeared as was expected, and Fig. S5(c) depicts these results. Note that the x-axis represents pixel numbers. By comparatively analyzing the dimensions of the imaging window that appeared in the coregistered US images, we found that the pixel distance "40" corresponded roughly to the position of the outer surface of the imaging window with a 3.38-mm OD. Moreover, based on the approximate location of the peak point of the blue dashed line, which we portrayed as a gross trend of the PA signal distribution [Fig. S5(c)], we could estimate that the optical WD was located around the distance of 50 pixels. Consequently, since there was a ~10-pixel value difference between the positions of the imaging window's outer surface and the optical WD, we concluded that the laser beam's focal spot traced a circle with a diameter measuring 3.54 mm, as depicted in Fig. S5(d), which was slightly different from the result (3.6 mm) calculated from Fig. S1.

Furthermore, when looking at the image in Fig. S5(c), how is it that a colorectal PA signal is detected at a position that is even more toward the inside of the 3.38-mm diameter outer surface of the imaging window? We attribute the result to the off-axial positioning (i.e., fluctuation) of the scanning tip with respect to the central axis of the imaging window during scanning [see Fig. 5(e) and Visualization 3] (note that the two radial dimension values included in Fig. S5(c) were the measures determined with respect to the center of the scanning tip, not from the tubular imaging window).